# Double metal-insulator transitions and MR in $La_{0.67}Ca_{0.33}Mn_{1-x}Ru_xO_3$ (x ≤ 0.10): A qualitative understanding in light of possible magnetic phase separation

L. Seetha Lakshmi, V. Sridharan[*], D.V. Natarajan, Rajeev Rawat[a], Sharat Chandra, V.Sankara Sastry and T.S. Radhakrishnan[b]

Materials Science Division, Indira Gandhi Centre for Atomic Research,
Kalpakkam 603102, Tamil Nadu, India
[a] Inter University Consortium for DAE Facilities, Khandwa Road, Indore,
Madhya Pradesh, India.
[b]Present contact e-address: tsr_res@vsnl.net

## Abstract:

We report results of the magnetotransport measurements on $La_{0.67}Ca_{0.33}Mn_{1-x}Ru_xO_3$ (0 ≤ x ≤ 0.1) compounds in the light of proposed magnetic phase separation in our previous work wherein two metal to insulator transitions (MITs) were reported (*Ref.: L.Seetha Lakshmi et.al, J.Magn.Magn. Mater. 257, 195 (2003)*). With the application of magnetic field, a significant reduction in resistance and a shift in MITs to higher temperatures but with different rates are observed. The temperature dependent magnetoresistance (MR) exhibits two maxima, a pronounced one at higher temperature and a shallow one at low temperature, corresponding to the two MITs. The peak value of MR at high temperature MITs (H=5T) does not change systematically with Ru concentration. The low temperature MR peak becomes perceptible for x ≥ 0.03 and shows a nominal enhancement with Ru concentration. The double maxima, both exhibiting magnetoresistance phenomena in the magnetotransport properties of Ru doped manganites are analysed within the context of magnetic phase separation.

PAC numbers: 75.30 .Vn

-------------------------------------------------------------------------------------------------------------------

[*]Corresponding Author: Dr. V. Sridharan,
                    Materials Science Division,
                    Indira Gandhi Centre for Atomic Research,
                    Kalpakkam, Tamil Nadu – 603 102, India
                Email: sridh61@hotmail.com
                Phone: +91-04114-280081: Fax-091-04114-280081

**Introduction:**

The underlying mechanism governing the colossal magnetoresistance (CMR) in the mixed valent perovskite manganites is still a matter of discussion. Enormous efforts in the theoretical as well as in the experimental fronts have clearly established that the earlier proposals, such as Double Exchange Mechanism [1-3], strong electron phonon coupling manifesting in the form of Jahn–Teller polarons [4-6], Anderson type localizations [7-8] are inadequate for a realistic explanation for the CMR phenomena and its ground state properties. The detailed numerical studies by Dagotto *et. al* [9-12] indicate that manganites, belonging to strongly correlated electron systems exhibit a tendency towards intrinsically inhomogeneous states. Abundant experimental works also have established the presence of such inhomogeneous states [13-19]. Thus, the CMR phenomenon is shown to be linked to phase co-existence over a wide range of temperature with different magnetic and electrical properties and to the competition between them. From the present understanding, the mixed phase tendencies are broadly categorised into two types: the *electronic phase separation* involving two phases with different electronic density with nanometer scale clusters and the *disorder induced phase separation* with percolative characteristics between equal density of states near its first order MITs forming co-existing clusters in the sub-micrometer scale [13].

As the essential degrees of freedom pertinent to the manganites such as spin, charge, lattice and orbital are intimately linked to Mn ion, any perturbation at the Mn site is expected to modify its ground state properties [20-26]. In this context, role of Mn-site substitutions on the electrical as well as the magnetotransport properties should be studied within the framework of mixed phase tendency. In the process of understanding the role of local structure due to the modified ionic size at the Mn site as well as the local magnetic coupling on the ground state properties of CMR manganites, we have carried out diamagnetic ($Al^{3+}$, $Ga^{3+}$, $Ti^{4+}$, $Zr^{4+}$ and $Hf^{4+}$) as well as paramagnetic substitutions ($Fe^{3+}$ and $Ru^{4+}$) at the Mn site of $La_{0.67}Ca_{0.33}MnO_3$, a canonical example of intermediate bandwidth CMR manganites [27-30]. It is found that all substitutions suppress the transition temperatures to different extent and eventually modify the Ferromagnetic-Metallic (FM-M) ground state to a cluster glass (CG). Wide variation in the rate of suppression of the transition temperatures on substitution with a FM-M ground state are understood in terms of the effect of local structure as well as local magnetic coupling at the Mn site. It is noteworthy that paramagnetic $Ru^{4+}$ ion, with smaller ionic radius as compared to $Mn^{3+}$ suppresses the transition temperatures at a rate of 3 K/at. %, the least reported in the Mn site substituted CMR manganites [29]. Among the paramagnetic substitutions, Ru is reported to be the most efficient in inducing ferromagnetism and metallicity in the charge ordered rare earth manganites [31-35], with a MIT as high as 240 K, where as in others it was reported to be typically less than 150K [32]. Hence, Ru appears to be exceptional among the paramagnetic substitutions, which will provide a better insight into the role of local magnetic coupling on the transport properties of manganites.

In our previous work [29], we have reported the structural, electrical and magnetic properties of Ru doped $La_{0.67}Ca_{0.33}MnO_3$ compounds. From the increase in the lattice parameters and the expansion in the unit cell in the composition range ($x \leq$ 0.085) studied, we infer Ru to be in the mixed valence state of $Ru^{3+}$ and $Ru^{4+}$, rather than $Ru^{4+}$ and/or $Ru^{5+}$ reported in other manganites [31,35]. Double MITs - *characterised by a high temperature maximum followed by a broad low temperature* *maximum spanning a relatively wider temperature range, both shifting to lower temperatures, at a rate of 2.3 and 17 K/at% respectively-* could be seen in the temperature dependence of electrical resistivity. Unlike the high temperature maximum, no magnetic feature could be discerned either in ac susceptibility or in dc magnetization associated with the low temperature broad maximum in the electrical resistivity. Based on the structural, magnetic and electrical properties, we proposed a magnetic phase separation in the system with two FM-M phases coexisting in the ground state. In this paper, we report the magnetotransport properties of $La_{0.67}Ca_{0.33}Mn_{1-x}Ru_xO_3$ ($0 \leq x \leq 0.10$) compounds under an applied magnetic field of 1 and 5T. The field dependent resistance was also measured at 300, 235, 125 and 5 K. The experimental results will be analysed within the framework of a *magnetic phase separation* occurring in the system.

**Experiment:**

$La_{0.67}Ca_{0.33}Mn_{1-x}Ru_xO_3$ ($0 \leq x \leq 0.10$) compounds were synthesized by a standard solid state reaction between the stoichiometric mixtures of $La_2O_3$, $CaCO_3$, $MnO_2$ and $RuO_2$ according to the procedure reported in our previous work [29]. High-resolution room temperature powder X-ray diffraction (XRD) patterns were recorded in the $2\theta$ range 15-120° using a STOE (Germany) diffractometer. The temperature variation of resistivity in the temperature range 300 - 4.2 K was measured in van der Pauw geometry [36]. Silver paint was used for the electrical contacts for the disc shaped pellets. The ac susceptibility measurements were performed on powder samples using a home-built ac susceptometer under an average field of 0.25 Oe at a frequency of 947 Hz. In our discussion, the in-phase component of ac susceptibility ($\chi'$) alone is considered. The magnetoresistance measurements (MR) in an applied magnetic field H of 1 and 5T were carried out on the rectangular bar shaped pellets by a standard four-probe method. A superconducting magnet was employed with the magnetic field direction parallel to the current direction. The field dependent MR measurements were also carried out at 300, 235, 120 and 5 K. Differential Scanning Calorimeter (Mettler DSC821[e]) measurements were performed in the temperature interval 160 -300 K with Ar as purge gas.

**Results and Discussion**

The room temperature powder XRD patterns of the compounds are shown in Fig. 1. From the XRD studies, the compounds were found to be monophasic. Over the entire range of composition of Ru under the present study, the structure was determined to be orthorhombic (SG: P$nma$). The lattice parameters and the unit cell volume were found to increase with Ru substitution. The detailed analysis of the structural characterization was published elsewhere [29]. The temperature variation of resistivity ($\rho(T)$) of La$_{0.67}$Ca$_{0.33}$Mn$_{1-x}$Ru$_x$O$_3$ (0 $\leq$ x $\leq$ 0.10) compounds in the absence of an applied magnetic field are shown in Fig. 2. The virgin compound, La$_{0.67}$Ca$_{0.33}$MnO$_3$ undergoes a metal to insulator transition (MIT) characterised by a maximum at T=T$_{MI}$ in the $\rho(T)$ curve. Close to this transition, a ferromagnetic to paramagnetic transition at T=T$_C$ could also be observed. Interestingly, Ru substitution results in two maxima in the $\rho(T)$ curve, a high temperature maximum at T=T$_{MI1}$ followed by a low temperature maximum at T=T$_{MI2}$ spanning relatively wider temperature range. Both the maxima were found to shift to lower temperatures with the Ru substitution. While the former shifts at a rate of ~3K/at. % up to x=0.085 and much more rapidly for x=0.10, the smallest rate reported for the Mn-site substituted CMR manganites, the latter shifts at a rate of 16 K/at. % (Fig. 3). This is in contrast to charge ordered Ru doped $La_{0.4}Ca_{0.6}MnO_3$ [33], wherein the second peak is reported not to shift with doping but remained centred at 125 K.

The ferro to paramagnetic transition temperature closely matches with $T_{MI1}$ and decreases with Ru substitution at the same rate as that of $T_{MI1}$. But no characteristic feature of magnetic transition associated with $T_{MI2}$ could be observed in the $\chi'$ signal [29]. Figure 4 shows the normalised DSC thermograms of compounds in the temperature interval of 160 to 300 K. A smooth background (dashed line in Fig. 4) common to these thermograms was subtracted and the resultant DSC curves are shown as the lower right inset in Fig. 4. In all the thermograms, a single peak alone could be seen. This peak shifts to lower temperatures (upper inset panel of Fig. 4) and becomes progressively broader with an increase of Ru concentration. These observations are in general agreement with that of susceptibility measurements. The progressive broadening of DSC peaks as well as $\chi'$ signal with an increase in Ru doping clearly demonstrates that the system is magnetically inhomogeneous. This is further supported by a *non-liner decrease* in the magnetic moment upon Ru substitution [29].

In our previous work, we have interpreted the double MITs in the Ru doped $La_{0.67}Ca_{0.33}MnO_3$ compounds in terms of possible magnetic phase separation: two ferromagnetic metallic (FM-M) phases co-existing in the ground state. The FM-M phase with $dT_C/dx\sim3K/at.$ % corresponds to a $Ru^{4+}$ rich region wherein the $Ru^{4+}$ couples *ferromagnetically* with the neighbouring Mn spins. The other FM-M phase (with $dT_C/dx\sim16$ K/at.%) being a $Ru^{3+}$ rich region with a *local antiferromagnetic* coupling of the $Ru^{3+}$ with neighbouring Mn spins. Of the two, the latter is a poor conducting magnetic phase. Weigand et. al. from MXCD measurements, have shown such an local antiferromagnetic coupling between mixed valent $Ru^{3+}$ and $Ru^{4+}$ and Mn ions for the Ru substituted layered manganite $La_{1.2}Sr_{1.8}Mn_2O_7$ [37].The presence of a local antiferromagnetic coupling between $Fe^{3+}$ (isoelectronic of $Ru^{3+}$) and neighbouring Mn ions in $La_{0.67}Ca_{0.33}Mn_{1-x}Fe_xO_3$ is well-established [38].

Complementing the previous work, magnetoresistance as a function of temperature under the applied magnetic field of 1 and 5T were carried out. For the sake of clarity, representative curves are shown in Fig. 5. With the application of magnetic field, both the maxima broaden and shift to higher temperatures (Fig. 5), a characteristic feature of the CMR manganites. As the magnetic field further broadens the inherently broad $T_{MI2}$, we could not determine the same for x = 0.01 and 0.03. While the magnetic field of 1T shifts $T_{MI1}$ more or less constantly by 14K (Fig. 6), for 5T the peak shifts beyond the maximum temperature range of measurement, viz., 300K . On the other hand, shift of $T_{MI2}$ under 1 and 5T is *not constant* over the doping range of present study; it decreases with an increase of Ru composition and for x = 0.10, a negligible shift is observed (Fig. 6).

A significant reduction in the resistance, on application of magnetic field was observed over the entire temperature range of study. The magnetoresistance defined as

$$MR = -100* (R[H]-R [0])/ R [0]$$

was calculated and are shown in Fig. 7 (a) and (b) for H=1 and 5T respectively. It is seen from Fig. 7(a) that for all Ru concentrations, the MR (H=5T) exhibits a dominant maximum at high temperature. A weak but definite maximum at lower temperature is also observed (marked by arrows) which becomes dominant and distinct for x = 0.1. There is a overall agreement between the $T_{MI}$ s and corresponding maximum in the MR. The magnetoresistance curves for H=1T exhibits a prominent maximum corresponding to $T_{MI1}$, a semblance of it could be seen at lower temperatures (corresponding to $T_{MI2}$). The high temperature MR peaks for H=5T are considerably broader compared to that of H=1T. Also, for a given magnetic field, the MR peak shifts to lower temperature, in accordance with the shift of $T_{MI1}$ and gets progressively broader as well with an increase of Ru concentration. There also exists a significant magnetoresistance away from the peaks showing a weak temperature dependence on further lowering the temperature, a characteristic feature of polycrystalline manganites [39,40].

Figure 8 (a, b, c & d), shows the variation of MR at 300K, 235K ($T_{MI1}$ of x = 0.1), 120 K (~ $T_{MI2}$ of x = 0.1) and 5K respectively. In the paramagnetic state (300 K), MR of the samples decreases with field exhibiting a $H^2$ dependence and at any given field strength, it decreases with the increase of Ru. In contrast to this, the low temperature MR exhibits a sharp linear decrease at low fields (<1T) followed by a much slower linear decrease above 1 T. The low field MR absent for the single crystals, is due to grain boundary (GB) arising from the intergrain tunnelling of the spin polarised electrons [41,42]. The GB MR enhances as the temperature is lowered leading to a well-developed knee structure (Fig. 8(c) and 8(d)). The MR of the x = 0.1 compound is the highest at 235 K and 120 K (Fig. 8(b) and 8(c)) which correspond to its $T_{MI1}$ and $T_{MI2}$. The contribution of GB to MR is estimated by back extrapolating the linear high field slope and finding its intercept at zero magnetic field [40]. The value of the MR due to GB at 5 K is ~ 24% and does not vary much with substitution.

Presence of double maxima feature in the $\rho(T)$ curve has been reported by many workers. Most of these reports had attributed the origin of the double maxima feature to extrinsic factors such as the grain size (GS), grain boundary (GB) and oxygen off-stiochiometry [43-49]. Presence of two peaks have also been reported for the Ln-site substituted $LnMnO_3$ system with $Ce^{4+}$ ion [43]. However, it was shown that samples were multiphasic evidenced by the presence of additional peak(s) in the XRD pattern. From the high statistics XRD pattern collected, we exclude the presence of impurity phase with volume fraction more than 1% and no peak was left unindexed. Additionally, the transition temperatures determined from the resistivity, ac susceptibility and DSC measurements are in close agreement and decrease linearly over the entire composition range ruling out impurity phase formation. Hence, secondary phase could be ruled out as being the cause for the double maxima in the $\rho(T)$. Two maxima in the $\rho(T)$ are reported for samples with GS less than 1 $\mu$m and the MR(T,H) curves exhibit a single peak only [49]. On the other hand, as the GS of all samples of present study are in the range 15 to 20 $\mu$m [29], we unambiguously exclude the GS as being the cause of low temperature maximum. The insensitiveness of the GBMR to the Ru substitution implies that GB is not altered an extent to warrant a low temperature maxima [49].

Oxygen deficiency ($\delta$) in $La_{0.7}Ca_{0.3}MnO_{3-\delta}$ is also reported to result in two maxima in the resistivity curve [49]. With the increase of $\delta$, the "magnitude" of the low temperature maximum increases and shifts to lower temperature. If the low temperature maximum observed in the present study is due to oxygen off-stoichiometry, then from the reported shift of this peak the value of $\delta$ is estimated to be as high as $\delta=0.2$. However, such a large value of $\delta$ is reported to produce a structural change [50], which we do not observe as evident from XRD data. Also, such large value of $\delta$ should have resulted in drastic increase in the value of $\rho_o$, typically by few orders of magnitude. On the other hand, such a large change in $\rho_o$ is not observed in our compounds. We conclude that oxygen off-stoichiometry is not the cause for the low temperature maximum upon Ru substitution. From the foregoing discussions, it is seen that the low temperature maximum in $\rho(T)$ curve is intrinsic and is associated with a metal to insulator transition. It is noteworthy that such a feature is present in certain paramagnetic ions substituted at the Mn site of CMR manganites [23,51,52].

The peak in the resistivity of (La:Ca)MnO$_3$ system is understood in the context of phase separation; coexistence of a ferromagnetic-metallic (FM-M) phase and an antiferromagnetic insulating phase. The FM-M phase, present as clusters of ~12Å size in the paramagnetic-insulating regime, grows as the temperature is lowered from the paramagnetic state [16]. As the critical volume ($V_C$) for percolative conductivity is reached, a sharp fall in the resistance is observed. This results in a maximum in the resistivity. At a given temperature, application of magnetic field enhances the volume fraction of the FM-M phase and the $V_C$ is attained at a higher temperature. This results in the shift of the maximum to higher temperature; higher the field strength, larger will be the shift of the maximum. Thus, the application of the magnetic field results in the enhancement in the volume fraction of FM-M phase leading to reduction in the resistivity and shifts the maximum to higher temperature. However, dramatic changes in the resistivity will be seen close to $V_C$, i.e. near $T_{MI}$, largely due to the contrast in the conductivity of the percolating conducting phase and the insulating matrix. The shift of the maximum to higher temperature along with reduction in $\rho$ in the presence of magnetic field results in colossal magnetoresistance at about $T_C$.

The scenario is somewhat different in the case of magnetic phase separation wherein two ferromagnetic metallic phases co-exist in its ground state. As the sample is cooled down from higher temperatures (i.e. paramagnetic insulating phase) a ferromagnetic phase (FM-M1) evolves within a paramagnetic (PM) matrix, having appreciable resistivity contrast between the phases. Upon reaching $V_C$, FM-M1 establishes a percolative path and the resistivity drops drastically resulting in a maximum in $\rho(T)$. On further cooling, the PM phase itself undergoes a para to ferromagnetic transition. As the system is percolatively conducting below the $T_{MI1}$, magnetic transition of the PM phase is not expected to produce drastic drop in the resistivity, as seen for $T_{MI1}$. The magnetic field results in an expected shift of $T_{MI1}$ to higher temperature. Such a shift is not expected in the case of $T_{MI2}$ for the reason that the magnetic transition of PM phase takes place within a conducting FM-M1 matrix with a low conductivity contrast. Such a low conductivity contrast results in an almost vanishing magnetoresistance maximum corresponding to $T_{MI1}$ for H = 1 T. This signal grows into a broad maximum under a magnetic field of 5 T but far smaller compared to that of $T_{MI1}$. These features, negligible shift of $T_{MI2}$ and substantially reduced MR maximum associated with $T_{MI2}$ support our conjecture of magnetic phase separation occurring in $La_{0.67}Ca_{0.33}Mn_{1-x}Ru_xO_3$.

In passing, we also wish to bring another interesting observation concerning the magnetoresistance. As pointed out, the higher temperature MR maximum, for H=1 & 5 T shifts to lower temperature with Ru doping (Fig. 6(a) & 6(b)). However, the value of MR at the maximum does not change appreciably among the Ru substituted systems. This is in contrast to La-site substituted systems wherein magnitude of MR is shown to exponentially increase with the decrease of $T_C$.

In summary, Ru substituted system exhibits two maxima in the resistivity variation with temperature. It is shown that they are intrinsic and are attributed to the metal to insulator transition. While the high temperature transition shifts to higher temperature with the application of magnetic field, relatively a negligible shift is observed for the low temperature transition. This result in an appreciable magnetoresistance associated with the former transition, but only a sparing magnetoresistance for the low temperature transition.  These features are explained in terms of magnetic phase separation: coexistence of two ferromagnetic metallic phases in its ground state.

 **Acknowledgement:**

 Authors thank Ms. T. Geetha Kumary and Dr. Y. Hariharan for extending the low temperature experimental facilities. L.Seetha Lakshmi also thanks Council of Scientific and Industrial Research, New Delhi for awarding a Senior Research Fellowship.

**References:**

[1]    C. Zener, Phys. Rev. 82 (1951) 403.

[2]    P.W. Anderson, H. Hasegawa, Phys. Rev. 100 (1955) 675.

[3]    P.G. de Gennes, Phys. Rev. 118 (1960) 141.

[4]    A. J. Millis, Nature 392 (1998) 147.

[5]    A. J. Millis, P.B. Littlewood, B.I. Shraiman, Phys. Rev. Lett. 74 (1995) 5144.

[6]    A. J. Millis, B.I. Shraiman, R. Mueller, Phys. Rev. Lett. 77 (1996) 175.

[7]    J. M. D. Coey, M. Viret, L. Ranno, K. Ounadjela, Phys. Rev. Lett. 75  (1995)
        3910.

[8]    L. Sheng, D.Y. Xing, D.N. Sheng, C. S. Ting, Phys. Rev. B 56 (1997) R 7053.

[9]    E. Dagotto, T. Hotta, A. Moreo,  Phys. Reports 344 (2001) 1.

[10]   J. Burgy, M. Mayor, V. Martin-Mayor, A. Moreo, E. Dagotto, Phys.  Rev. Lett.
        87 (2001)   277202.

[11]   A. Moreo, M. Mayor, A. Feiguin, S. Yunoki, E. Dagotto, Phys. Rev. Lett. 84
        (2000) 5568.

[12]   M. Mayor, A. Moreo, J. Verges, J. Arispe, A. Feiguin, E. Dagotto,
        Phys. Rev.Lett 86 (2001) 135.

[13]   M. Uehara, S. Mori, C. H. Chen, S. W. Cheong,  Nature 399 (1999) 560.

[14]   M. Fath, S. Freisem,  A. A. Menovsky, Y. Tomioka, J. Aarts , J. A. Mydosh,
        Science 285, (1999) 1540.

[15]   S. J. L.  Billinge, Th. Proffen, V. Petkov, J. L. Sarrao, S. Kycia,
        Phys. Rev. B 62 (2000-II) 1203.

[16]   J. M. De Teressa, M. R .Ibarra , P. A .Algarabel, C. Ritter, C. Marquina,
        J. Blasco, J. Garcia, A. del Moral, Z. Arnold, Nature 386  (1997) 256 .

[17]   J. W. Lynn, R. W. Erwin, J. A. Borchers, Q. Huang, A. Santoro, J. L Peng,

Z.Y. Li, Phys. Rev. Lett. 76 (1996) 4046.

[18]   G. Allodi, R. De Renzi, G. Guindi, F. Licci, M.W. Pieper,  Phys. Rev. B

56 (1997) 6036.

[19]   R. H. Heffner, J. E. Sonier, D. E. Mac Laughlin, G. J. Nieuwenhuys, G. Ehlers,

F. Mezei, S. W. Cheong, J. S. Gardner, H. Röder, Phys. Rev. Lett. 85 (2000)

3285.

[20]   J. Blasco, J. Garcia, J.M. de Teresa, M.R. Ibarra, J. Perez, P.A. Algarabel,

C. Marquina, C. Ritter, Phys. Rev. B  55 (1997) 8905.

[21]   Xianming Liu, Xiaojun Xu, Yuheng Zhang, Phys. Rev. B 62 (2000) 15 112.

[22]   Young Sun, Xiaojun Xu, Lei Zheng, Yuheng Zhang, Phys. Rev. B 60 (1999)

12317.

[23]   Young Sun, Xiaojun, Yuheng Zhang, Phys. Rev. B 63 (2001) 54404.

[24]   N. Gayathri, A.K. Raychaudhuri, S.K. Tiwary, R. Gundakaram, Anthony Arulraj,

C.N.R. Rao, Phys. Rev. B 56 (1997) 1345.

[25]   K. M. Krishnan, H. L. Ju, Phys. Rev. B 60 (1991) 14 793.

[26]   K.H. Ahn, X.W. Wu, K. Liu, C.L. Chien, J. Appl. Phys. 81 (1997)  5505.

[27]   V. Sridharan, L. Seetha Lakshmi, R. Nithya, R. Govindraj, D.V.Natarajan,

T.S.Radhakrishnan, J. Alloys. Comp. 326 (2001) 65.

[28]   L.Seetha Lakshmi, V. Sridharan, D.V. Natarajan, V. Sankara Sastry,

T.S. Radhakrishnan, Pramana, J. Phys. 58 (2002)1019.

[29]   L. Seetha Lakshmi, V. Sridharan, D.V. Natarajan, V. Sankara Sastry ,

T.S. Radhakrishnan, Ponnpandian, R. Justine Joseyphus,

A. Narayananasamy, J. Magn. Magn. Mater,  257  (2003) 195.

[30]   L.Seetha Lakshmi, V. Sridharan, D.V. Natarajan, Sharat Chandra,

R.Rajaraman, V. Sankara Sastry ,T.S. Radhakrishnan ( under communication).

[31]  A. Maignan, C. Martin, M. Hervieu, B. Raveau, J. Appl. Phys. 89 (2001) 500.

[32]  S. Hebert, A. Maignan, C. Martin, B. Raveau, Solid State Commun. 121 (2002) 229.

[33]  B. Raveau, A. Maignan, C. Martin, R. Mahindran, M. Hervieu, J. Solid State Chem. 151 (2000) 330.

[34]  J. S. Kim, D. C. Kim, G. C. McIntosh, S. W. Chu, Y. W. Park, B. J. Kim, Y. C. Kim, A. Maignan, B. Raveau, Phys. Rev. B 66 (2002) 224427.

[35]  C. Autret, C. Martin, A. Maignan, M. Hervieau, B. Raveau, G. Andre, F. Bouree, A. Kurbakov V. Trounov, J. Magn. Magn. Mater. 241(2002) 303.

[36]  van der Pauw, Philips Res. Repts, 13 (1958) 1.

[37]  F. Weigand, S. Gold, A. Schmid, J. Geissler, E. Goering, K. Dorr, G. Krabbes, R. Ruck, Appl. Phys. Lett. 81 (2002) 2035.

[38]  S. B. Ogale, R. Shreekala, R. Bethe, S. K. Date, S.I. Patil, B. Hannoyer, F. Petit, G. Marest, Phys. Rev. B 57 (1998) 7841.

[39]  H.L. Ju, J. Gopalakrishnan, J.L. Peng, Q. Li, G.C. Xiong, T. Venkatesan, R.L. Greene, Phys. Rev. B 51 (1995) 6143.

[40]  H.L. Ju, H. Sohn, Solid State Commun.102 (1997) 463.

[41]  H. Y. Hwang, S. W. Cheong, N. P. Ong, B. Batlogg, Phys.Rev.Lett 77 (1996) 2041.

[42]  P. Lyu, D.Y. Xing, J. Dong, J. Magn. Magn. Mater. 202 (1999) 405.

[43]  J. R. Gebhardi, S. Roy, N. Ali, J. Appl. Phys. 85 ( 1999) 5390.

[44]  N. Zhang, W. Ding, W. Zhong, D. Xing, Y. Du, Phys. Rev. B 56 ( 1997) 8138.

[45]  P. Mandal, S. Das, Phys. Rev. B 56 (1997) 15703.

[46]  M. Izumi, Y. Konishi, T. Nishihara, Appl. Phys. Lett. 73 ( 1998) 2497.

[47]  J. R. Sun, G. H. Rao, Y. Z. Zhang, Appl. Phys. Lett. 72 (1998) 3208.

[48]  K. M. A. Hossain,  L. F. Cohen, T.  Kodenkandeth, J. Macmanus- Driscoll,
       N. McN. Alford, J. Magn. Magn.Mater. 195 (1999) 31.

[49]  L. Malavasi, M. C. Mozzati, C. B. Azzoni, G. Chiodelli, G. Flor, Solid State
       Commun.123 ( 2002) 321.

[50]  C. Ritter, M. R. Ibarra, J. M. De Teresa, P. A. Algarabel , C. Marquina,
        J. Blasco, J. Garcia, S. Oseroff, S. W. Cheong,  Phys. Rev. B 56 (1997) 8902.

[51]  C. L. Yuan, Y. Zhu, P. P. Ong, Solid State Commun.120 (2001) 495.

[52]  F. Rivadulla, M. A. Lopez- Quintela, L. E. Hueso, P. Sande, J. Rivas,
       R. D. Sanchez, Phys. Rev. B 62 (2000) 5678.

**Figure Captions:**

1.  High statistics room temperature powder X-ray diffraction patterns of
    $La_{0.67}Ca_{0.33}Mn_{1-x}Ru_xO_3$ compounds ( x = 0, 0.01, 0.03, 0.05, 0.07, 0.085 & 0.10 ).

2.  Temperature dependence of resistivity of $La_{0.67}Ca_{0.33}Mn_{1-x}Ru_xO_3$ compounds
    ( x = 0, 0.01, 0.03, 0.05, 0.07, 0.085 & 0.10) in the absence of magnetic field.

3.  Variation of metal to insulator transition temperatures ($T_{IM1}$ & $T_{IM2}$) as a function of
    Ru concentration (x) of $La_{0.67}Ca_{0.33}Mn_{1-x}Ru_xO_3$ compounds ( x = 0, 0.01, 0.03,
    0.05, 0.07, 0.085 & 0.10).  Dotted lines are the best linear fit to the experimental
    data to estimate the rate of suppression in the transition temperatures.

4.  DSC curves of $La_{0.67}Ca_{0.33}Mn_{1-x}Ru_xO_3$ compounds ( x = 0, 0.01, 0.03, 0.05, 0.07,
    0.085 & 0.10).  Upper panel of the inset shows the variation of the peak
    temperature as a function of Ru concentration. Dotted line is the best linear fit to
    estimate the rate of suppression of the transition temperature. Lower panel of the
    inset shows the variation of the area under the peak as a function of Ru
    concentration, after subtracting a common background. Refer text for details.

5.  Temperature dependence of resistance of $La_{0.67}Ca_{0.33}Mn_{1-x}Ru_xO_3$ compounds.
    Under a magnetic field of 0  (closed circles), 1 (solid line) and 5 T (dotted line).
    Representative curves of x = 0,0.03, 0.05, 0.07 & 0.10 are given for the sake of
    clarity.   Refer text for details.

6.  Variation of magnetoresistance peaks  ($T_{IM1}$ & $T_{IM2}$) as a function of Ru
    concentration of $La_{0.67}Ca_{0.33}Mn_{1-x}Ru_xO_3$ compounds ( x = 0,  0.03, 0.05, 0.07 &
    0.10 ) under a magnetic field of  1 and 5T.

7.  Temperature dependence of magnetoresistance (MR) of $La_{0.67}Ca_{0.33}Mn_{1-x}Ru_xO_3$
    compounds. Representative curves of x = 0.03, 0.05, 0.07 & 0.10 are given. Up

(a) H= 1T; (b) H= 5T. For the sake of clarity,  similar curves  for x=0.0 are given in the inset of each panel

8.  Field dependence of magnetoresistance at  300 (a),  235 (b), 120 (c)  &  5 K (d) of $La_{0.67}Ca_{0.33}Mn_{1-x}Ru_xO_3$ compounds (x =0, 0.03, 0.05, 0.07 & 0.10).

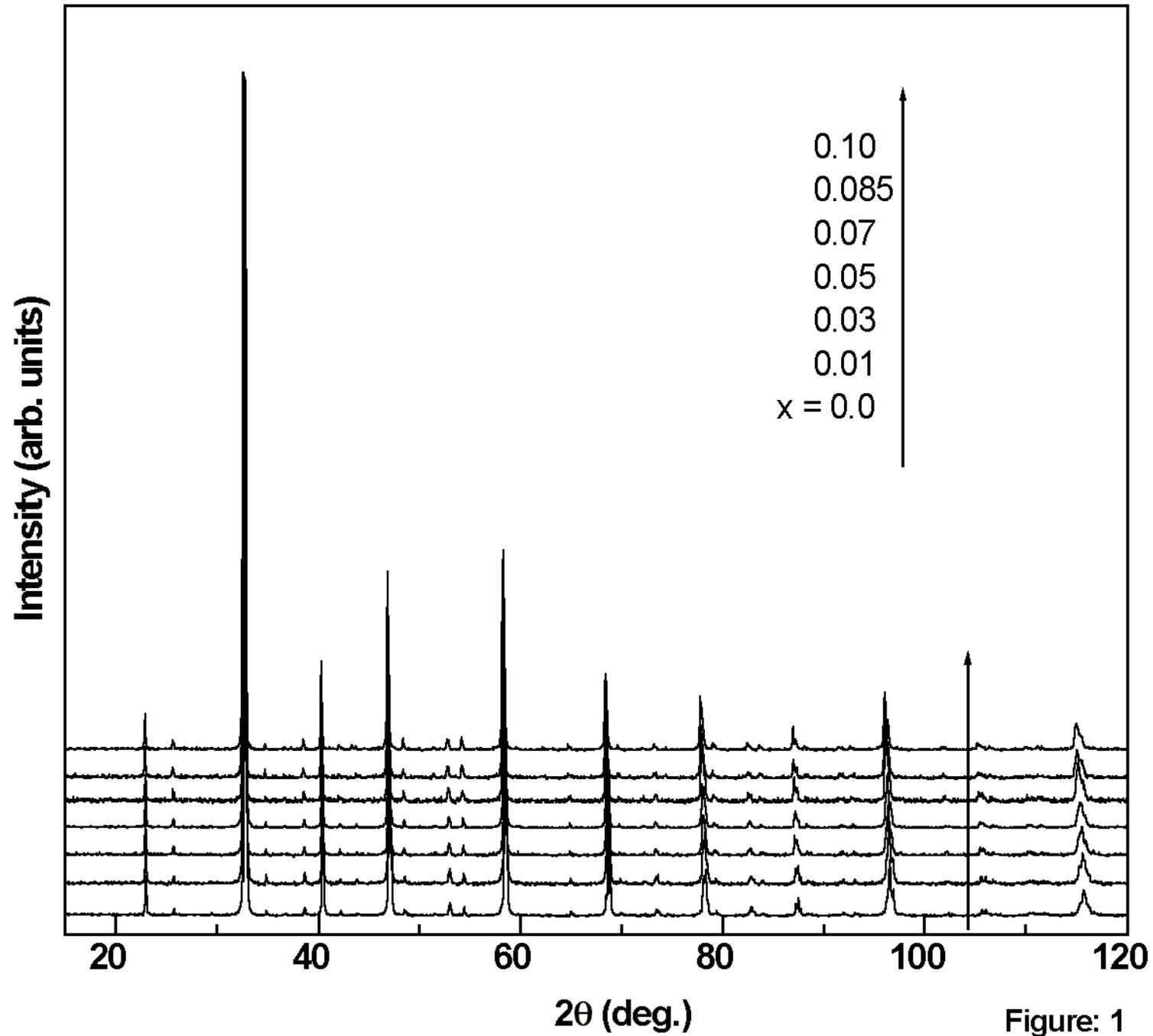

0.10
0.085
0.07
0.05
0.03
0.01
x = 0.0

**Intensity (arb. units)**

**2θ (deg.)**

Figure: 1

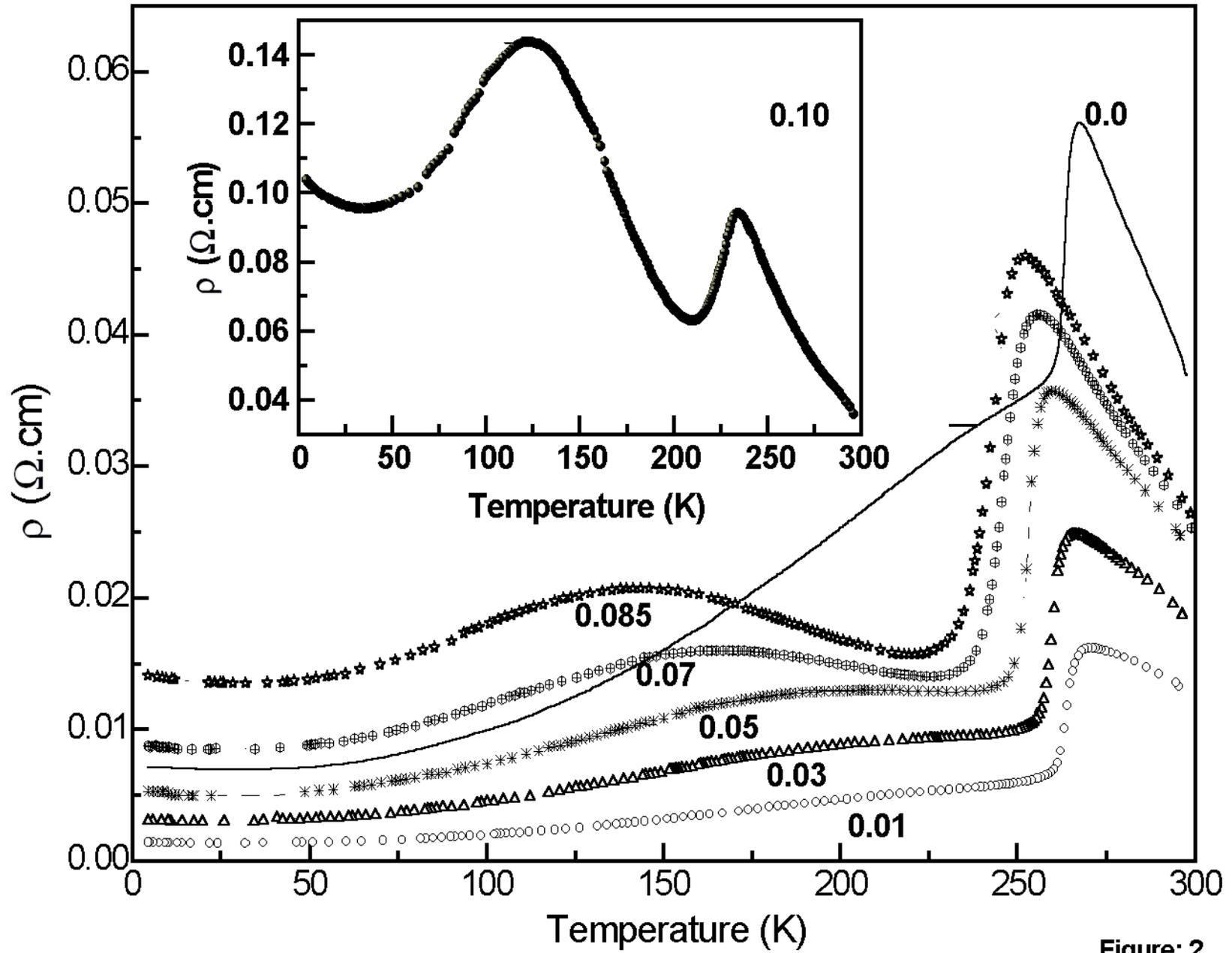

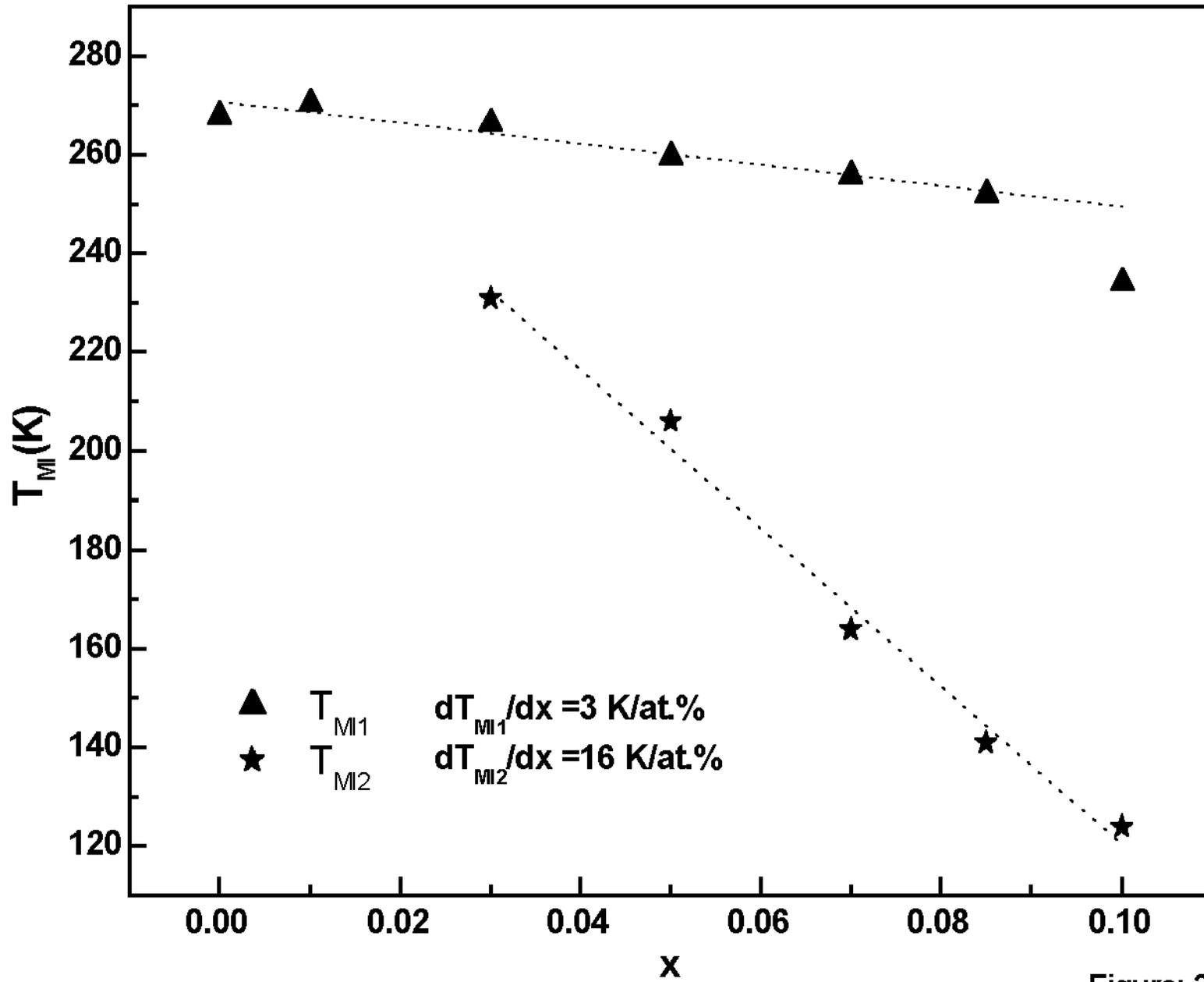

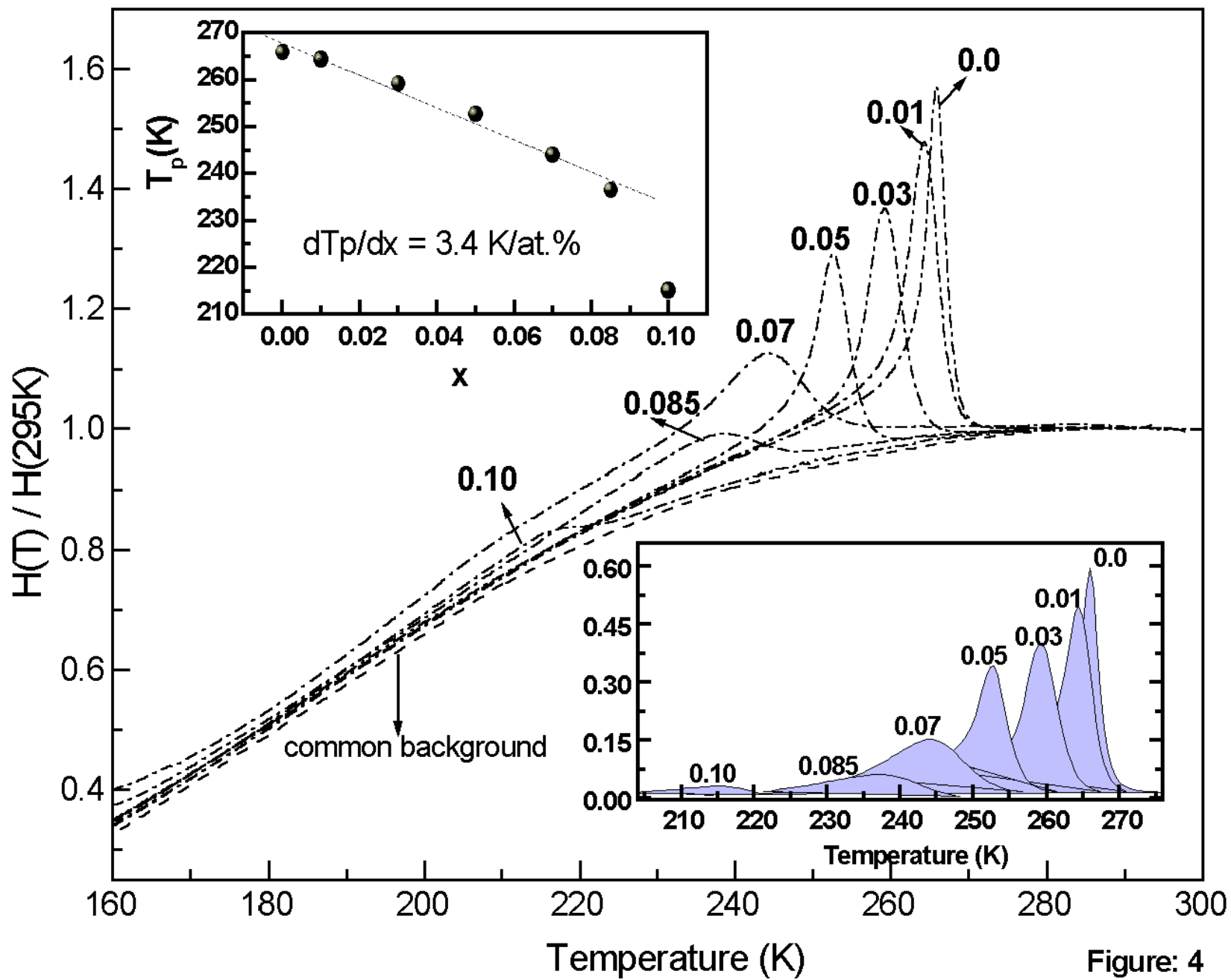

Figure: 4

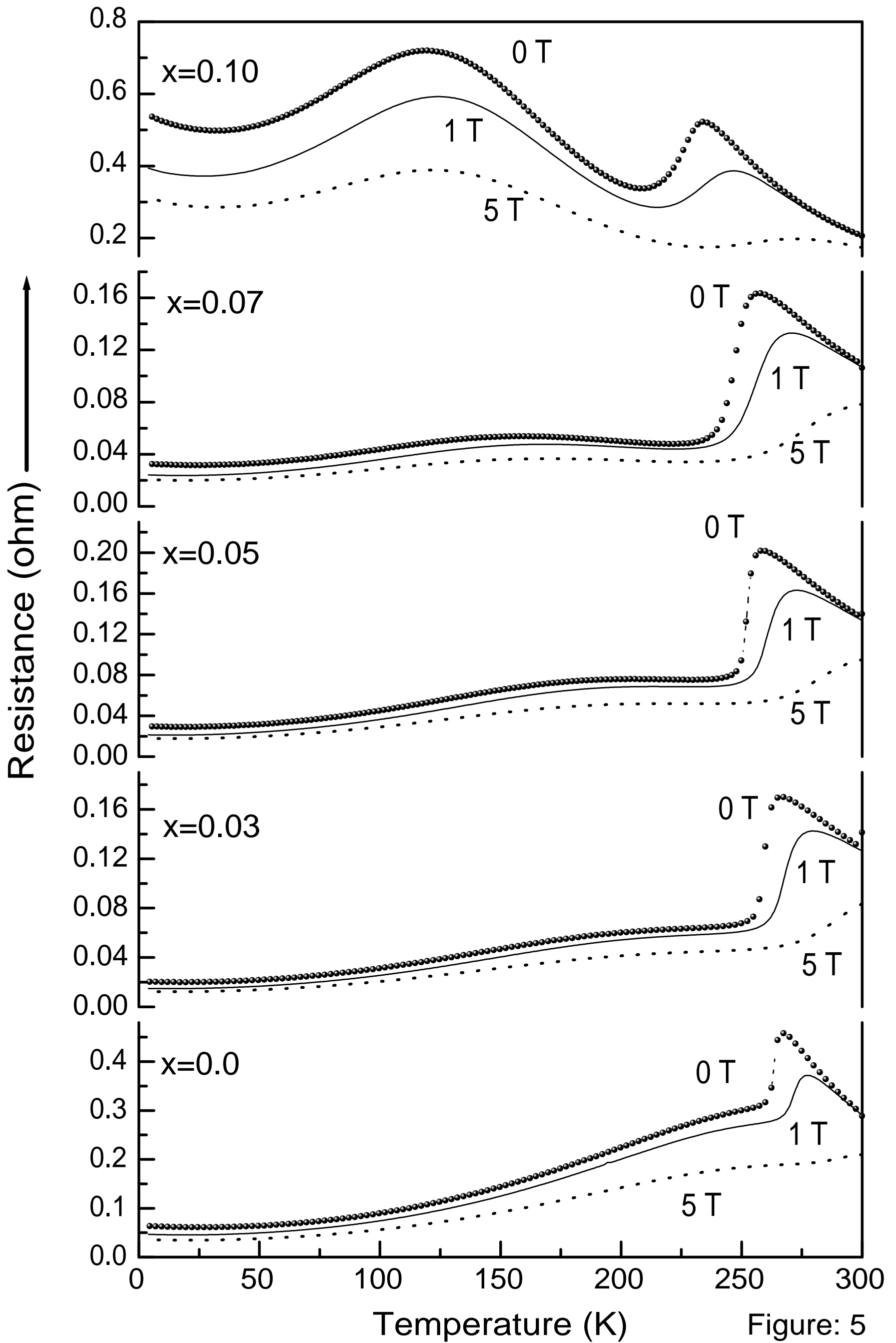

Seetha Lakshmi et. al.

Figure: 5

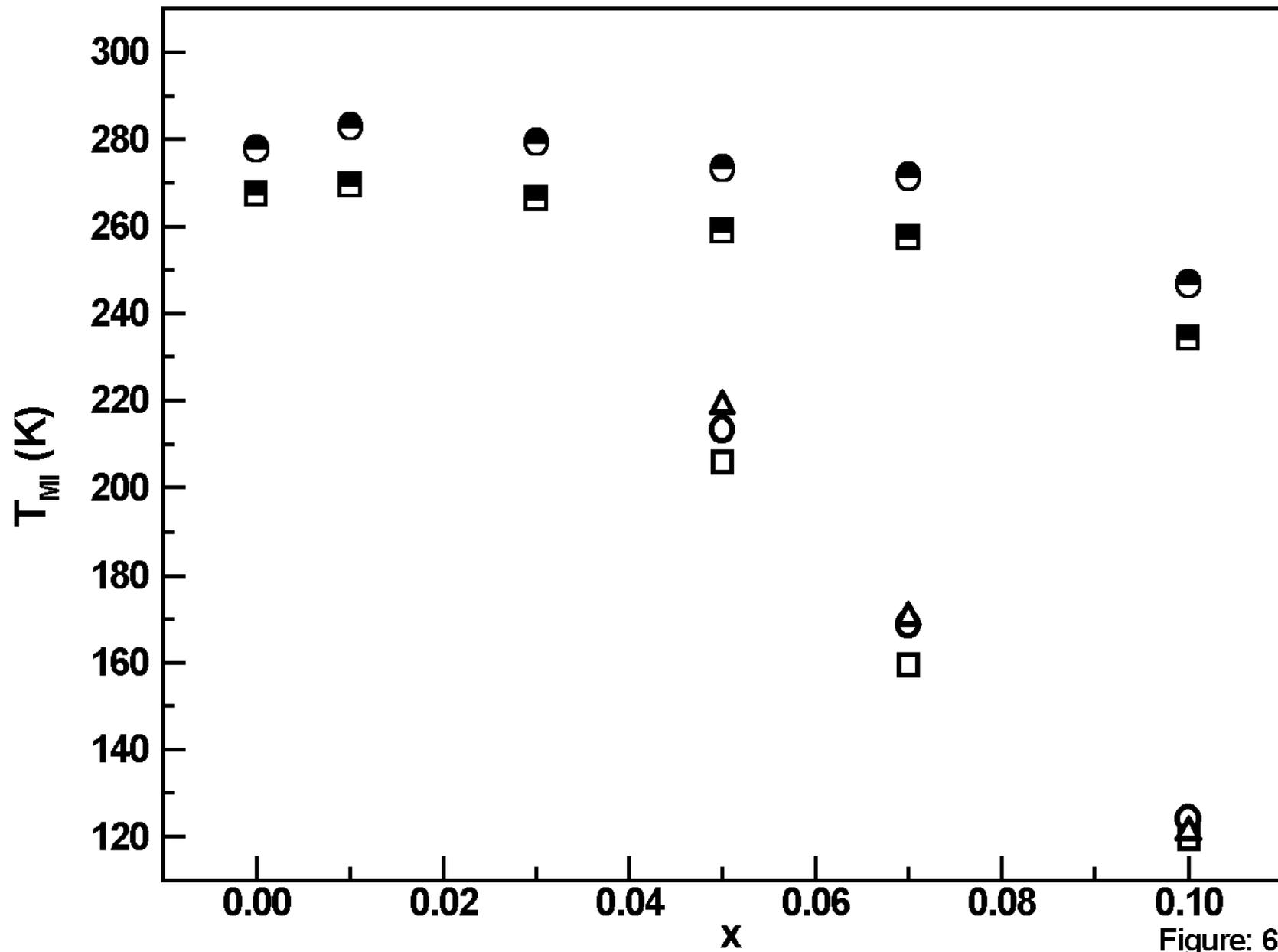

Figure: 6

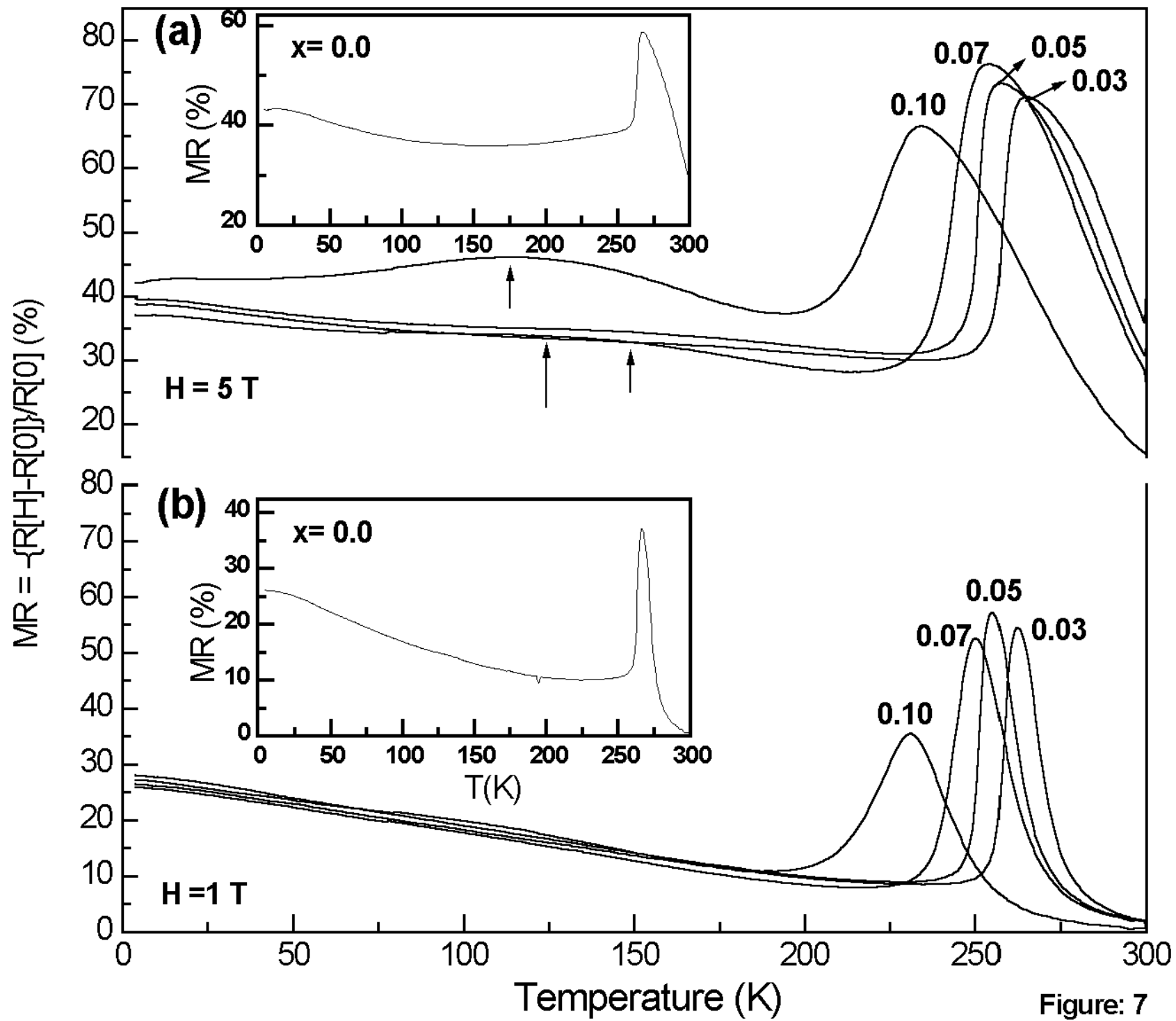

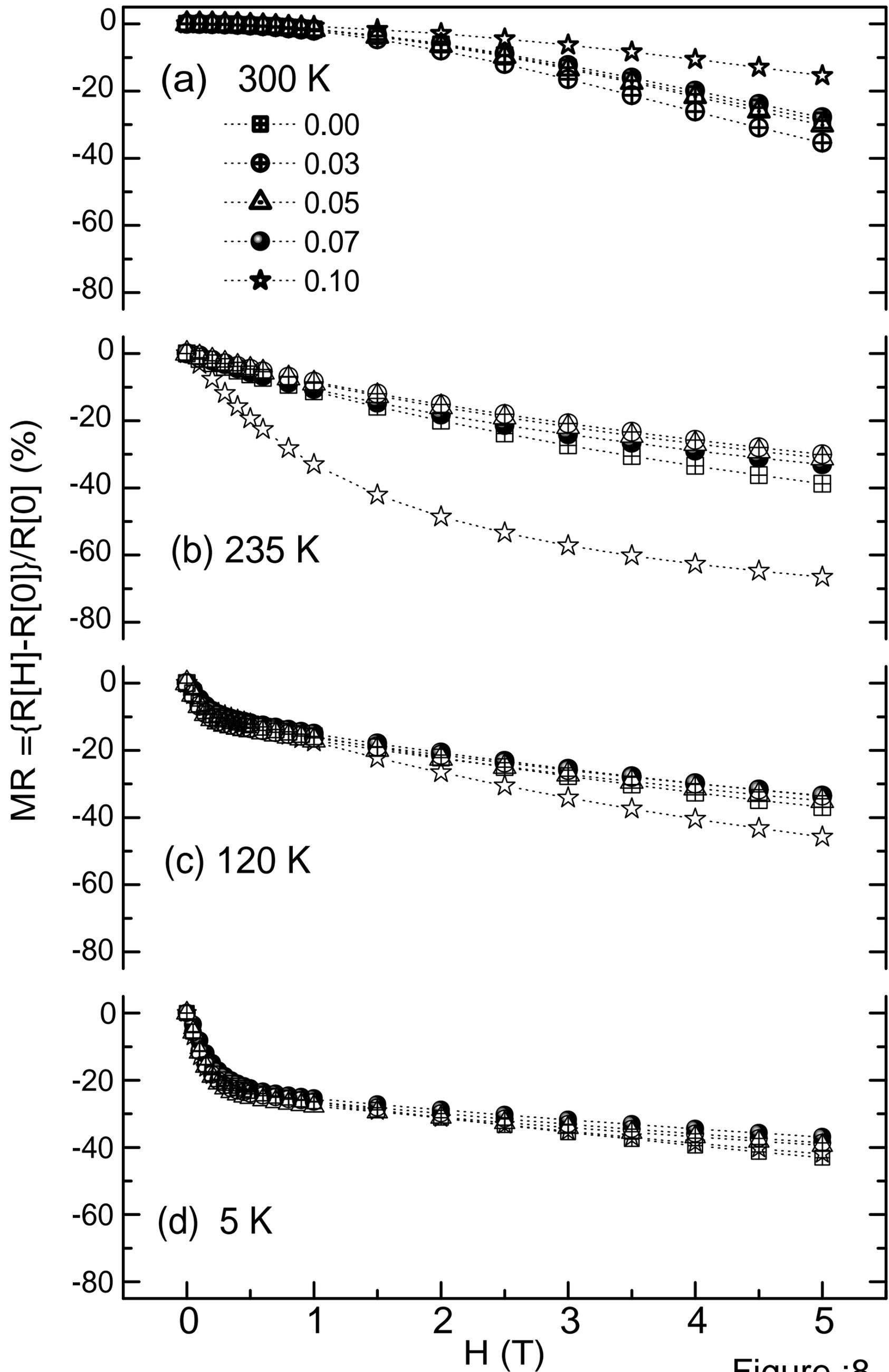

Seetha Lakshmi et. al

(a) 300 K

- ⊞ 0.00
- ⊕ 0.03
- △ 0.05
- ● 0.07
- ☆ 0.10

(b) 235 K

(c) 120 K

(d) 5 K

MR = {R[H]−R[0]}/R[0] (%)

H (T)

Figure : 8